\documentclass[lettersize,journal]{IEEEtran}

\IEEEoverridecommandlockouts
\usepackage{algorithm} 		
\usepackage{algpseudocode}
\usepackage{cite}
\usepackage{amsmath,amssymb,amsfonts}
\usepackage{float}
\usepackage{graphicx}
\usepackage{textcomp}
\usepackage{xcolor}
\usepackage{color}
\usepackage{stfloats}
\usepackage{subcaption}
\usepackage{array}
\usepackage{multirow}
\usepackage{caption}
\usepackage{subcaption}
\usepackage{booktabs}
\usepackage{tabularx}
\usepackage[colorlinks=true,linkcolor=black,citecolor=colorlinks=true,linkcolor=black,citecolor=blue,urlcolor=blue,urlcolor=blue]{hyperref}

\begin{document}

\title{\LARGE Error Rate Analysis and Low-Complexity Receiver Design for Zero-Padded AFDM}
\author{Qin Yi,~\IEEEmembership{Graduate~Student~Member,~IEEE,} Zeping Sui,~\IEEEmembership{Member,~IEEE,} and Zilong Liu,~\IEEEmembership{Senior~Member,~IEEE}
\vspace{-2em}

\thanks{

Qin Yi is with the National Key Laboratory of Wireless Communications, University of Electronic Science and Technology of China 611731, Sichuan, China (e-mail: yiqin@std.uestc.edu.cn).

Zeping Sui and Zilong Liu are with the School of Computer Science and Electronics Engineering, University of Essex, Colchester CO4 3SQ, U.K. (e-mail: zepingsui@outlook.com, zilong.liu@essex.ac.uk).  
\emph{(Corresponding author: Zeping Sui.)}

}
}

\maketitle
\begin{abstract}This paper studies the error rate performance and low-complexity receiver design for zero-padded affine frequency division multiplexing (ZP-AFDM) systems. By exploiting the unique ZP-aided lower triangular structure of the time domain (TD) channel matrix, we propose a novel low-complexity minimum mean square error (MMSE) detector and a maximum ratio combining-based TD (MRC-TD) detector. Furthermore, the theoretical bit error rate (BER) performance of both the MMSE and maximum likelihood detectors {is} analyzed. Simulation results demonstrate that the proposed detectors can achieve identical BER performance to that of the conventional MMSE detector based on matrix inversion while enjoying significantly reduced complexity.
\end{abstract}

\section{Introduction}
The next generation of wireless communication systems is {deemed} to support reliable data transmission in high-mobility scenarios, including high-speed {trains}, vehicle-to-everything systems, and {low-earth-orbit satellite} networks \cite{HighMob}. {In those scenarios, the conventional orthogonal frequency division multiplexing (OFDM) may experience significant performance degradation due to Doppler-induced inter-carrier interference (ICI) \cite{OFDM_Disadvan}}\footnote{{Under frequency selective Rayleigh fading channels, Doppler shifts in OFDM can be estimated and cancelled using maximum likelihood (ML) techniques \cite{TurboCoded_R1Add_2,OFDM_R1Add_1}. However, in doubly selective channels, residual Doppler estimation errors may remain \cite{OFDM_R1Add_2}, which motivates the investigation of Doppler-resilient waveforms.}}. Against this background, affine frequency division multiplexing (AFDM) \cite{AFDM} has emerged recently as a promising Doppler-resilient waveform. In AFDM, {every information symbol} is modulated using the inverse discrete affine Fourier transform (IDAFT) {onto} {a chirp subcarrier that occupies} the entire bandwidth, thereby {leading to enhanced} robustness against Doppler spread. {At the same time, AFDM enjoys excellent backwards compatibility as it can be efficiently implemented through the minimal modification of the OFDM systems \cite{Chirp}. Additionally,} AFDM is capable of achieving a separable channel representation and attaining full diversity in doubly selective channels by appropriately tuning the chirp rate according to the maximum Doppler spread \cite{AFDM_Para1}. 

{There has been a substantial body of research on AFDM, including pilot-aided channel estimation for AFDM \cite{AFDM}, non-orthogonal AFDM \cite{NonAFDM_1,NonAFDM_2}, AFDM aided integrated sensing and communications \cite{AFDM_ISAC}, as well as various integration schemes with sparse code multiple access \cite{AFDM_SCMA}, index modulation \cite{AFDM_IM}, and generalized spatial modulation \cite{AFDM_GSM}}. However, most existing studies {have only} considered AFDM with a cyclic periodic prefix (CPP). As a parallel development, zero padding (ZP) has been widely adopted in OFDM as an alternative to cyclic prefix (CP) to eliminate inter-symbol interference (ISI) while preserving the advantageous sparsity and lower-triangular structure of the time-domain (TD) channel matrix \cite{ZP-OFDM,ZP-OFDM1,ZP-OFDM2}.
{Moreover, ZP-based multicarrier systems enjoy higher power transmission efficiency, ensure reliable symbol recovery under deep channel fading (i.e., channel spectral nulls), and provide enhanced capabilities in channel estimation and tracking \cite{ZP-OFDM}.}
These structural and performance benefits motivate us to exploit ZP-{aided} AFDM systems {in this work}.

{Although low-complexity linear detection schemes, such as the maximum-ratio combining (MRC) detector, have been developed for CPP-AFDM systems by exploiting the approximate sparsity of the DAFT-domain channel matrix \cite{AFDM}, they suffer from noticeable performance degradation compared with the conventional {minimum mean square error (MMSE) detector}. Our work aims to address this backdrop.}
The main contributions are summarized as follows:
\begin{itemize}
\item We derive the generic transceiver architecture for the proposed ZP-AFDM system, and design two low-complexity detectors: 1) MMSE detector based on Cholesky decomposition {without matrix inversion}, and 2) MRC-based TD (MRC-TD) detector, followed by detailed complexity analysis. Both detectors leverage the sparse and lower-triangular structure of the TD channel matrix {inherent to ZP-AFDM systems.}

\item {We derive the closed-form bit error rate (BER) expressions for ZP-AFDM under both {ML} and MMSE detection. For the ML detector, the conditional pairwise error probability (PEP) is first obtained, followed by the derivation of a tight BER upper bound based on the union bound technique. For the MMSE detector, the BER is approximated by analyzing the signal-to-interference-plus-noise ratio (SINR) of each chirp subcarrier.}
\item {Numerical results demonstrate that 1) the derived BER curves closely match the results from Monte Carlo simulations; 2) the proposed ZP-AFDM outperforms its CPP-AFDM counterpart, as a higher power will be allocated to the data symbols for the former; and 3) the proposed detectors attain {BER} performance nearly identical to that of the conventional MMSE detector while offering significantly reduced complexity.}
\end{itemize}

\section{System Model}\label{Sec:SystemModel}
\begin{figure}
\centering
\includegraphics[width=0.9\linewidth] 
{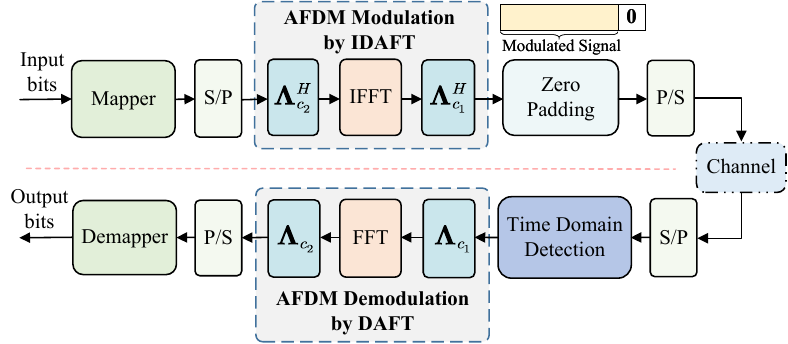}
\caption[left]{Transceiver diagram of the proposed ZP-AFDM system.}
\label{fig:ZPAFDM_SystemModel}
\end{figure}

As demonstrated in Fig.~\ref{fig:ZPAFDM_SystemModel}, the information symbol vector $\mathbf{x}=[x[0],x[1],\ldots,x[N-1]]\in \mathbb{C}^{N \times 1}$ in the affine frequency domain is transformed into the TD by using IDAFT, yielding
\begin{align}
\label{eq:Modu_Sym}
    {s}[n]=\frac{1}{\sqrt{N}}\sum\limits_{m=0}^{N-1} {x}[m] e^{\imath2\pi\left(c_1n^2+c_2m^2+\frac{nm}{N}\right)}, 
\end{align} 
for $n=0, ..., N-1$, where $N$ denotes the number of chirp subcarriers, and $c_1$ and $c_2$ represent the chirp parameters. The TD signal {in} \eqref{eq:Modu_Sym} can be rewritten in matrix form as $\mathbf{s} = \mathbf{A}^H\mathbf{x}$, where $\mathbf{A} = \mathbf{\Lambda}_{c_2} \mathbf{F} \mathbf{\Lambda}_{c_1}$ represents the DAFT matrix,  $\mathbf{\Lambda}_{c} = \text{diag}\left(e^{-\imath 2\pi c n^2}, \, n = 0, \ldots, N-1 \right)$ and $\mathbf{F}$ is the unitary discrete Fourier matrix with entries $\frac{1}{\sqrt{N}} e^{-\imath \frac{2\pi}{N} m n}$. To mitigate ISI, {an} $N_{\mathrm{ZP}}$-length ZP is appended to the modulated signal, where $N_{\mathrm{ZP}}$ is no smaller than the maximum normalized delay {spread}.


We consider a doubly selective channel with $P$ propagation paths, whose impulse response at time $n$ and delay $l$ is expressed as
\begin{align}\label{eq:channel}
g_n(l) = \sum_{i=1}^{P} h_i \, e^{-\imath \frac{2\pi}{N} \nu_i n} \, \delta(l - l_i),
\end{align}
where $h_i$, $\nu_i \in [-\nu_{\max}, \nu_{\max}]$, and $l_i$ denote the complex-valued path gain, normalized Doppler shift, and integer delay of the $i$-th path, respectively, and $\delta(\cdot)$ is the Dirac delta function. The maximum normalized Doppler shift is denoted by $\nu_{\max}$, while the maximum delay spread is defined as $Q=\max(\mathcal{L})$, where $\mathcal{L}=\{l_1,\ldots,l_P\}$ denotes the set of normalized delay {indices}.

\begin{figure}
    \centering   
    \includegraphics[width=0.8\linewidth] {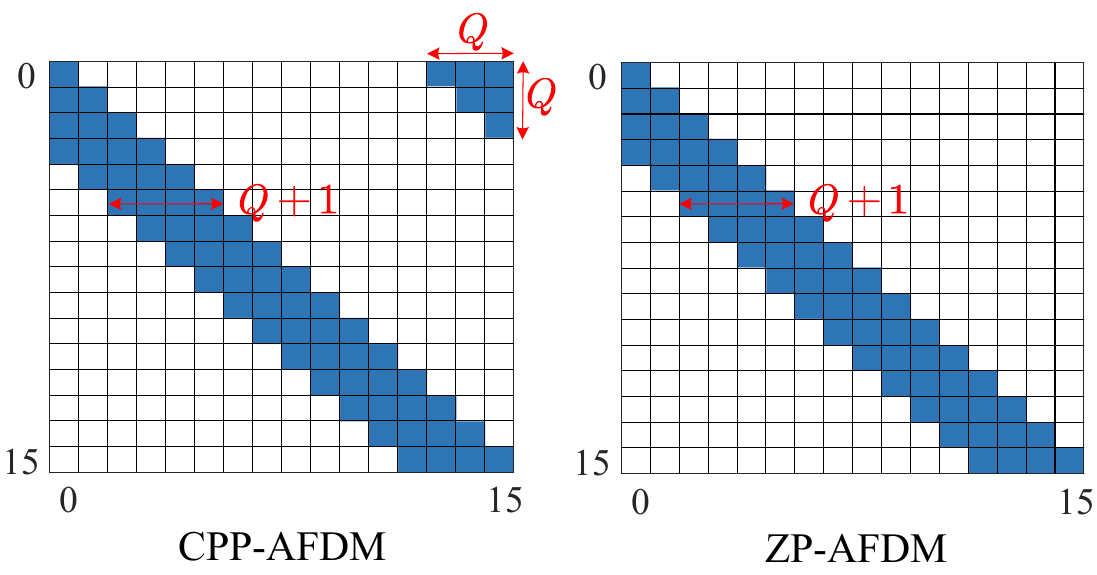}
    \caption{Illustration of the TD channel matrix $\mathbf{H}$ for CPP-AFDM and ZP-AFDM with $N=16$ and $Q = 3$.}
\label{fig:H_Tx}
\end{figure}

At the receiver, the discrete-time received signal can be expressed as
\begin{align}\label{eq:RxSym_TD}
  r[n] = \sum_{l\in\mathcal{L}} s[n-l] g_n(l) + w_\mathrm{T}[n],
\end{align}
{where $r[n]$ represents the $n$-th sample of the received TD signal, and $w_\mathrm{T}[n]\sim \mathcal{CN}(0, \sigma^2)$ denotes the additive white Gaussian noise. By stacking all received samples, the received TD signal can be equivalently expressed in vector form as ${\mathbf{r}} =  [r[0], r[1], \ldots, r[N-1]]^{T} = {\mathbf{H}} {\mathbf{s}} + {\mathbf{w}}_\mathrm{T}$}, where the TD channel matrix is given by ${\mathbf{H}} = \sum_{i=1}^{P} h_i {\mathbf{\Delta}}_{\nu_i} \boldsymbol{\Pi}^{l_i}$ with ${\mathbf{\Delta}}_{\nu_i} \triangleq \operatorname{diag}\left(e^{-\imath \frac{2 \pi}{N} \nu_i n}, n = 0, 1, \ldots, N-1\right)$ and {the forward linear shift matrix} $\boldsymbol{\Pi}$. 
{As illustrated in Fig.~\ref{fig:H_Tx}, $\mathbf{H}$ in ZP-AFDM exhibits a lower-triangular structure with a bandwidth of ($Q+1$), owing to the linear convolution enabled by ZP. In contrast, in the CPP-AFDM system, CPP transforms the channel convolution into a circular convolution, which introduces additional non-zero entries in the upper-right corner of $\mathbf{H}$.}
Upon utilizing DAFT, the affine frequency domain received signal can be expressed as ${\mathbf{y}} = {\mathbf A} {\mathbf{r}}  ={\mathbf H}_\mathrm{eff} {{\bf{x}}} + {\mathbf{w}}_\mathrm{AF}$, where $\mathbf{H}_\mathrm{eff} = \mathbf{A} \mathbf{H} \mathbf{A}^H$ denotes the effective channel matrix and $\mathbf{w}_\mathrm{AF} = \mathbf{A} \mathbf{w}_\mathrm{T}$ has the same statistics as $\mathbf{w}_{\mathrm{T}}$ since $\mathbf{A}$ is unitary. Consequently, the input-output relationship of our ZP-AFDM can be derived as 
\begin{align}\label{eq:RxSignal_AFD2}
{\mathbf{y}}  =\sum_{i=1}^{P} h_i {\mathbf{H}}_i \mathbf{x} + {\mathbf{w}}_\mathrm{AF},
\end{align}
where $\mathbf{H}_i = \mathbf{A} \mathbf{\Delta}_{\nu_i} \boldsymbol{\Pi}^{l_i} \mathbf{A}^H$ denotes the subchannel matrix of the $i$-th path, whose $(p,q)$-th entry is given by
\begin{align}\label{eq:SubHeff_Hi}
{H}_i(p,q)
=\frac{1}{N}
e^{{\imath}\frac{2\pi}{N}\left [
Nc_1l_i^2-\,q\,l_i
+N c_2(q^2-p^2)\right ]}\zeta (l_i,\nu_i,p,q),
\end{align}
with $\zeta (l_i,\nu_i,p,q)=\sum_{n=l_i}^{N-1} e^{-\imath \frac{2\pi}{N} (p - q + \nu_i + 2N c_1 l_i) n}$.

\section{Detection Algorithms for ZP-AFDM}
In this section, the ML detector is first presented, followed by the conventional MMSE detector. Furthermore, we propose low-complexity MMSE and MRC-TD detectors by exploiting the {sparse and} lower-triangular structure of the TD channel matrix. Finally, the complexity of the detectors is analyzed.

\subsection{ML Detection}
The optimal ML detector is expressed as
\begin{align}
\hat{\mathbf{x}}_{\text{ML}} = \arg\min_{\mathbf{x} \in \mathcal{A}^{N}} \left\| \mathbf{y} - \mathbf{H_\mathrm{eff}} \mathbf{x} \right\|_2^2,
\end{align}
where $\mathcal{A}$ denotes the modulation alphabet.

\subsection{Conventional MMSE Detection}
The linear MMSE detector in the affine frequency domain is written as
\begin{align}\label{eq:MMSE-AF}
\hat{\mathbf{x}}_{\text{MMSE}} = \mathbf{G}_\mathrm{AF} \mathbf{y} = \mathbf{G}_\mathrm{AF} \mathbf{H}_\mathrm{eff} \mathbf{x} + \mathbf{G}_\mathrm{AF} \mathbf{w}_\mathrm{AF},
\end{align}
where $\mathbf{G}_\mathrm{AF}$ denotes the equalization matrix in the affine frequency domain, yielding
\begin{align}
\mathbf{G}_\mathrm{AF} = \left(\mathbf{H}_\mathrm{eff}^H \mathbf{H}_\mathrm{eff} + \frac{1}{\gamma_s} \mathbf{I}_{N} \right)^{-1}\mathbf{H}_\mathrm{eff}^{H},
\end{align}
and $\gamma_s$ represents the signal-to-noise ratio (SNR) per symbol.



\subsection{Proposed Low-Complexity MMSE Detection}\label{Sec:CholeskyMMSE}
\begin{figure}
    \centering
    \includegraphics[width=0.98\linewidth] {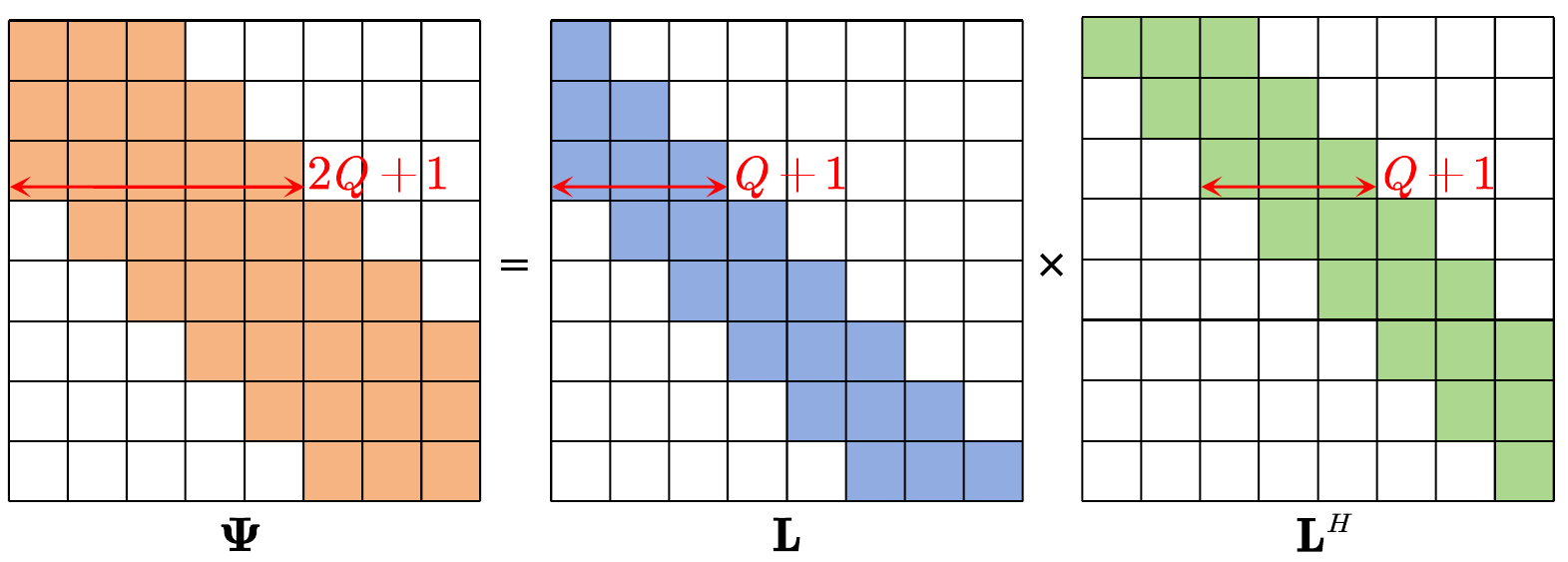}
    \caption[left]{The Cholesky factorization of the banded matrix $\mathbf{\Psi}$ with $N=8$ and $Q=2$.}
    \label{fig:CholeskyDecomp}
\end{figure}

{By utilizing $\mathbf{H}_\mathrm{eff} = \mathbf{A} \mathbf{H} \mathbf{A}^H$ and ${\mathbf{y}} = {\mathbf A} {\mathbf{r}}$, the MMSE detector in \eqref{eq:MMSE-AF} can be reformulated in TD as
\begin{align}\label{eq:MMSE-T}
\hat{\mathbf{x}}_{\text{MMSE}} 
&= \left ( \mathbf{A}\mathbf{H}^{{H}}\mathbf{H}\mathbf{A}^{\mathrm{H}} + \frac{1}{\gamma_s} \mathbf{I}_N \right )^{-1} \mathbf{A}\mathbf{H}^{{H}}\mathbf{r}\nonumber \\
&=\mathbf{A}\left ( \mathbf{H}^{{H}}\mathbf{H} + \frac{1}{\gamma_s} \mathbf{I}_N \right )^{-1} \mathbf{H}^{{H}}\mathbf{r}= \mathbf{A} \mathbf{\Psi}^{-1}\mathbf{H}^{H} \mathbf{r},
\end{align}
where $\mathbf{\Psi} = \mathbf{H}^{H} \mathbf{H} + \frac{1}{\gamma_s} \mathbf{I}_N$.}
Note that $\mathbf{H}^{H} \mathbf{H}$ is Hermitian and positive semi-definite, and the regularization term $\frac{1}{\gamma_s} \mathbf{I}_N$ is strictly positive for finite SNR, yielding the Hermitian and positive definite properties of $\mathbf{\Psi}$.
{Furthermore, as shown in Fig.~\ref{fig:CholeskyDecomp}, $\mathbf{\Psi}$ exhibits a banded structure with a bandwidth of ($2Q + 1$) since $\mathbf{H}$ is lower-triangular banded with a bandwidth of ($Q + 1$). By leveraging this structured sparsity, we apply the Cholesky decomposition \cite{Cholesky} to derive $\mathbf{\Psi} = \mathbf{L} \mathbf{L}^{H}$, where $\mathbf{L}$ remains lower-triangular banded with  a bandwidth of ($Q+1$).}
Then, the estimated symbol vector can be expressed as
\begin{align}\label{eq:MMSE-T1}
{\hat {\mathbf{x}}_{{\rm{MMSE}}}} = {\mathbf{A}}\underbrace {{{({{\mathbf{L}}^H})}^{ - 1}}\underbrace {{{\mathbf{L}}^{ - 1}}\overbrace {{{\mathbf{H}}^H}{\mathbf{r}}}^{\mathbf{b}}}_{\mathbf{z}}}_{{\mathbf{\hat s}}},
\end{align}
where $\mathbf{z} = \mathbf{L}^{-1} \mathbf{b}$ and $\hat{\mathbf{s}} = (\mathbf{L}^{H})^{-1} \mathbf{z}$ can be efficiently computed via forward and backward substitution, i.e., $\mathbf{L} \mathbf{z} = \mathbf{b}$ and $\mathbf{L}^H \hat{\mathbf{s}} = \mathbf{z}$, respectively. The detailed procedure is summarized in 
\textbf{Algorithm~\ref{alg:cholesky_mmse}}.

\begin{algorithm}
\caption{Proposed Low-Complexity MMSE Detection}
\label{alg:cholesky_mmse}
\begin{algorithmic}[1]
\footnotesize
\Require {TD} channel matrix \( \mathbf{H} \), {TD} received signal \( \mathbf{r} \), and SNR \(\gamma_s\)
\Ensure Estimated symbol vector \( \hat{\mathbf{x}} \)
\State \textbf{Notation:} In all loops, define \( j_\mathrm{start} = \max(0, i - Q) \), \( j_\mathrm{end} = \min(N-1, i + Q) \)
\State \textbf{Step 1: Compute} \( \mathbf{\Psi} = \mathbf{H}^H \mathbf{H} + \frac{1}{\gamma_s} \mathbf{I}_N \)

\For{$i = 0$ to $N-1$}
    \For{$j = j_{\text{start}}$ to $i$}
        \State $\mathit{\Psi } (i,j) = \mathbf{H}(:,i)^H \mathbf{H}(:,j)$
        \State $\mathit{\Psi } (j,i) = {\mathit{\Psi } (i,j)}^{*}$ 
    \EndFor
    \State $\mathit{\Psi } (i,i) = \mathit{\Psi } (i,i) + \frac{1}{\gamma_s}$
\EndFor

\State \textbf{Step 2: Cholesky Factorization} \( \mathbf{\Psi} = \mathbf{L} \mathbf{L}^H \)
\State $L(0,0) = \sqrt{\mathit{\Psi } (0,0)}$
\For{$i = 1$ to $N-1$}
    \For{$j = j_{\text{start}}$ to $i$}
        \If{$j = i$}
            \State $L(i,i) = \sqrt{\mathit{\Psi } (i,i) - \sum_{p=j_{\text{start}}}^{i-1} |L(i,p)|^2}$
        \Else
            \State $L(i,j) = \frac{\mathit{\Psi } (i,j) - \sum_{p=j_{\text{start}}}^{j-1} L(i,p) {L(j,p)}^*}{L(j,j)}$
        \EndIf
    \EndFor
\EndFor

\State \textbf{Step 3: Compute} $\mathbf{b} =\mathbf{H}^H \mathbf{r}$
\For{$i = 0$ to $N-1$}
    \State $b(i) = \sum_{p=i}^{j_{\text{end}}} H(p,i)^*\cdot r(p)$
\EndFor

\State \textbf{Step 4: Forward Substitution} \( \mathbf{L} \mathbf{z} = \mathbf{b} \)
\State $z(0) = \frac{b(0)}{L(0,0)} $
\For{$i = 1$ to $N-1$}
    \State $z(i) = \frac{1}{L(i,i)} \left[b(i) - \sum_{j=j_{\text{start}}}^{i-1} L(i,j) z(j) \right]$
\EndFor

\State \textbf{Step 5: Backward Substitution} \( \mathbf{L}^H \hat{\mathbf{s}} = \mathbf{z} \)
\State $\hat{s}(N-1) = \frac{z(N-1)}{L(N-1,N-1)^*} $
\For{$i = N-2$ to $0$}
    \State $\hat{s}(i) = \frac{1}{L(i,i)^*} \left[z(i) - \sum_{j=i+1}^{j_{\text{end}}} {L(j,i)^*} \cdot \hat{s}(j) \right]$
\EndFor
\State \Return $\hat{\mathbf{x}}={\mathbf{A}}\hat{\mathbf{s}}$.
\end{algorithmic}
\end{algorithm}

\subsection{Proposed MRC-TD Detection}\label{Sec:MRC-TD}
In this subsection, we propose an MRC-TD detector that directly exploits the sparse lower-triangular structure of the TD channel matrix $\mathbf{H}$. 
By coherently combining the nonzero channel coefficients, the proposed detector refines the desired symbol estimate while iteratively cancelling  ISI from previously detected symbols.

During the $k$-th iteration, each symbol $\hat{s}^{(k)}(n)$ is sequentially updated as
\begin{align}\label{Eq:sn}
\hat{s}^{(k)}{(n)} = \frac{g^{(k)}_n}{d_n + \frac{1}{\gamma_s}},
\end{align}
with
\begin{align}\label{Eq:g_k_n}
g_n^{(k)} = \sum_{p \in \mathcal{P}_n} H(p,n)^* \Delta r^{(k-1)}(p) + d_n \hat{s}^{(k-1)}(n),
\end{align}
\begin{align}\label{Eq:d_n}
d_n = \sum_{p \in \mathcal{P}_n} |H(p,n)|^2 ,
\end{align}
where $\Delta r^{(k-1)}$ denotes the residual error, and $\mathcal{P}_n = \{n+l \mid l\in\mathcal{L},\ n+l < N \}$ represents the set of {row indices corresponding to the non-zero entries} in column $n$ of matrix $\mathbf{H}$.
After each update, the residual vector $\Delta\mathbf{r}$ is refined by subtracting the contribution of the updated symbol as
\begin{align}\label{Eq:Delta_r}
\Delta r^{(k-1)}(p) \leftarrow \Delta r^{(k-1)}(p) - H(p,n)\left[ \hat{s}^{(k)}(n) - \hat{s}^{(k-1)}(n) \right],
\end{align}
for $p\in \mathcal{P}_n$. Once all the symbols are estimated, the estimated results will be exploited for interference cancellation in the next iteration. This process iterates among all the symbols until the updated input symbol vector is nearly identical to the previous one or the maximum iteration number $K$ is reached. Finally, the {estimated} symbol vector in the affine frequency domain is obtained as $\hat{\mathbf{x}} = \mathbf{A} \hat{\mathbf{s}}^{(K)}$. The MRC-TD detector is detailed in \textbf{Algorithm~\ref{alg:MRC-TD}}.


\textit{\textbf{Remark 1:} {Without loss of generality}, the proposed low-complexity MMSE and MRC-TD detectors for AFDM can be applied to OFDM by setting $c_1=c_2=0$.}

\begin{algorithm}
\caption{Proposed MRC-TD Detection}
\label{alg:MRC-TD}
\begin{algorithmic}[1]
\footnotesize
\Require {TD} channel matrix \( \mathbf{H} \), {TD received signal} \( \mathbf{r} \), maximum iterations \(K\), and SNR \(\gamma_s\)
\Ensure Estimated symbol vector \( \hat{\mathbf{x}} \)
\State Initialize \( \hat{\mathbf{s}}^{(0)} = \mathbf{0} \), \( \Delta \mathbf{r}^{(0)}= \mathbf{r} \)
\For{$k = 1$ to $K$}
    \For  {$n = 0$ to $N-1$}
        \State \( \mathcal{P}_n=\{ n + l \mid l \in \mathcal{L},\ n+l < N \}\)
        \State \( d_n = \sum_{p \in \mathcal{P}_n} |H(p,n)|^2 \)
        \State $g^{(k)}_n = \sum_{p \in \mathcal{P}_n}H(p,n)^*\Delta r^{(k-1)}(p) + d_n \hat{s}^{(k-1)}{(n)}$
        \State $\hat{s}^{(k)}{(n)} = \frac{g^{(k)}_n}{d_n + 1/\gamma_s}$
        \For {$p \in \mathcal{P}_n$}
           \State$\begin{aligned}
           &\Delta r^{(k-1)}(p) \leftarrow  \\ & \Delta r^{(k-1)}(p)- H(p,n)\left[\hat{s}^{(k)}{(n)} - \hat{s}^{(k-1)}{(n)}\right]
            \end{aligned}$
        \EndFor
    \EndFor
    \State$\Delta \mathbf {r}^{(k)}=\Delta \mathbf {r}^{(k-1)}$
    \State\textbf{if} {$\|\hat{\mathbf{s}}^{(k)}- \hat{\mathbf{s}}^{(k-1)}\|<\epsilon $} \textbf{then EXIT} 
\EndFor
\State \Return \( \hat{\mathbf{x}} = \mathbf{A} \hat{\mathbf{s}}^{(K)} \)
\end{algorithmic}
\end{algorithm}

\subsection{Complexity Analysis}
The conventional MMSE detector exhibits the complexity on the order of $\mathcal{O}(N^3)$. Based on the analysis in Subsection~\ref{Sec:CholeskyMMSE}, the computation {of $\mathbf{\Psi}$ and its Cholesky factorization are} on the order of $\mathcal{O}(NQ^2)$, while the complexity of forward and backward substitution is on the order of $\mathcal{O}(NQ)$, and we have $\mathcal{O}(N \log_2 N)$ for the complexity order of the final unitary transform. Therefore, the overall complexity of the proposed low-complexity MMSE detector is given by $\mathcal{O}(NQ^2 + N \log_2 N)$. For the proposed MRC-TD detector, since both symbol and residual updates involve at most $Q+1$ non-zero channel coefficients, the complexity per iteration can be formulated as $\mathcal{O}(NQ)$. Accordingly, considering $K$ iterations and the final unitary transform, the overall complexity of the proposed MRC-TD detector is on the order of $\mathcal{O}(KNQ + N\log_2 N)$.


\section{Performance Analysis}
\subsection{BER Analysis for ML Detection}
The received signal ${\mathbf{y}}$ in \eqref{eq:RxSignal_AFD2} can be rewritten as
\begin{align}
{\mathbf{y}}  =\sum_{i=1}^{P} h_i {\mathbf{H}}_i \mathbf{x} + {\mathbf{w}}_\mathrm{AF} = \mathbf{\Phi}(\mathbf{x}) \mathbf{h} +{\mathbf{w}}_\mathrm{AF},
\end{align}
where $\mathbf{h} = [h_1, h_2, ..., h_P]^T \in \mathbb{C}^{P \times 1}$ and
$\mathbf{\Phi}(\mathbf{x}) = [\mathbf{H}_1 \mathbf{x}, \mathbf{H}_2 \mathbf{x},\cdots, \mathbf{H}_P \mathbf{x}] \in \mathbb{C}^{N \times P}$. Assuming perfect channel state information at the receiver, the conditional PEP relying on the error event $(\mathbf{x} \rightarrow \hat{\mathbf{x}})$ can be expressed as \cite{PEP_R1Add}
\begin{align}\label{eq:PEP}
\mathrm{Pr}(\mathbf{x} \rightarrow \hat{\mathbf{x}} \mid \mathbf{h}) = Q \left( \sqrt{{\left \| {\mathbf \Phi}(\mathbf{\Delta }) \mathbf{h} \right \| ^2\cdot\gamma _s}/{2}} \right),
\end{align}
where $\mathbf{\Delta} =\mathbf{x}-\hat{\mathbf{x}}$.
Let ${\mathbf{\Theta}}={\mathbf \Phi}(\mathbf{\Delta })^H{\mathbf \Phi}(\mathbf{\Delta })$, which is a Hermitian matrix and can be eigenvalue decomposed as ${\mathbf{\Theta}} = {\mathbf{U\Sigma U^H}}$, where $\mathbf{U}$ is a unitary matrix and ${\mathbf{\Sigma}} = \mathrm{diag}\{\lambda_1, \lambda_2,\ldots, \lambda_P\}$. 
We then have ${\left \| {\mathbf \Phi}(\mathbf{\Delta }) \mathbf{h} \right \| ^2}=\mathbf{h}^H{\mathbf{\Theta}}\mathbf{h}=\mathbf{h}^H{\mathbf{U\Sigma U^H}}\mathbf{h}=\sum_{i=1}^{r} \lambda_i \left| \tilde{h}_i \right|^2$, where $r=\mathrm{rank}{({\mathbf{\Theta}})}$ and $\tilde{\mathbf{h}}=\mathbf{U}^H\mathbf{h}$. 
Applying $Q(x)\approx\frac{1}{12}\exp(-x^2/2)+\frac{1}{4}\exp(-2x^2/3)$, the conditional PEP in \eqref{eq:PEP} can be rewritten as 
\begin{align}
\mathrm{Pr}(\mathbf{x} \rightarrow \hat{\mathbf{x}} \mid \mathbf{h}) &\approx \frac{1}{12}\exp\left(-\frac{\gamma_s}{4}\sum_{i=1}^{r} \lambda_i 
\left|\tilde{h}_i\right|^2\right)\nonumber \\
&+ \frac{1}{4}\exp\left(-\frac{\gamma_s}{3}\sum_{i=1}^{r} \lambda_i \left|\tilde{h}_i\right|^2\right).
\end{align}
Assuming each path has an independent Rayleigh distribution, the unconditional PEP can be formulated as
\begin{align}\label{eq:UPEP}
\mathrm{Pr}(\mathbf{x} \rightarrow \hat{\mathbf{x}})\approx \frac{1}{12} \prod_{i=1}^{r} \frac{1}{1 + \frac{\lambda_i \gamma_s}{4P}} + \frac{1}{4} \prod_{i=1}^{r} \frac{1}{1 + \frac{\lambda_i \gamma_s}{3P}}.
\end{align}
{At high SNRs, \eqref{eq:UPEP} can be further approximated as
\begin{align}\label{eq:UPEP_HighSNR}
\mathrm{Pr}(\mathbf{x} \rightarrow \hat{\mathbf{x}})\approx \left(\frac{\gamma_s}{P}\right)^{-r}   \left(\frac{4^r}{12}+\frac{3^r}{4}\right) \prod_{i=1}^{r} \frac{1}{\lambda_i}.
\end{align}
It can be observed from \eqref{eq:UPEP_HighSNR} that the decay rate of the derived unconditional PEP in the high SNR regime is determined by \(r\), which also characterizes the diversity order \cite{AFDM}.}
Finally, the upper bound of the average BER is expressed as\footnote{{From \eqref{eq:SubHeff_Hi}, the entries of $\mathbf{H}_i$ explicitly depend on the AFDM-specific chirp parameters $c_1$ and $c_2$. Therefore, the key objects involved in \eqref{eq:PEP}--\eqref{eq:UPEP}, such as $\mathbf{\Phi}(\mathbf{\Delta})$, $\mathbf{\Theta}$, its rank $r$, and the eigenvalues ${\lambda_i}$, as well as the BER upper bound in \eqref{eq:UpperBound}, are all related to $c_1$ and $c_2$.}}
\begin{align}\label{eq:UpperBound}
P_e \leq \frac{1}{M^NN\log_2 M} \sum_{\mathbf{x}} \sum_{\substack{\mathbf{x} \neq \hat{\mathbf{x}}}} \mathrm{Pr}(\mathbf{x} \rightarrow \hat{\mathbf{x}}) e(\mathbf{x}, \hat{\mathbf{x}}),
\end{align}
where $e(\mathbf{x}, \hat{\mathbf{x}})$ denotes the number of bit errors between the symbol vectors $\mathbf{x}$ and $\hat{\mathbf{x}}$.

\subsection{BER Analysis for MMSE Detection}
{According to \eqref{eq:MMSE-AF} and \eqref{eq:MMSE-T}, the MMSE detectors in the affine frequency domain and TD are mathematically equivalent, yielding identical detection performance. For analytical convenience, the following performance analysis is based on the TD expression of the MMSE detector in \eqref{eq:MMSE-T}, which can be further written as
\begin{align}\label{eq:MMSE_T_Final}
\hat{\mathbf{x}}_{\mathrm{MMSE}}
= \mathbf{A}\mathbf{\Psi}^{-1}\mathbf{H}^{H}\mathbf{H}\mathbf{A}^{H}\mathbf{x}
+ \mathbf{A}\mathbf{\Psi}^{-1}\mathbf{H}^{H}\mathbf{w}_{\mathrm T}.
\end{align}
Hence, the $i$-th element of the detected signal $\hat{\mathbf{x}}_{\mathrm{MMSE}}$ can be expressed as
\begin{align}\label{eq:DetectedSignal_ith}
 \hat x(i) = T(i,i)x(i) + \sum\limits_{j \ne i} T (i,j)x(j) + \bar w(i),
\end{align}
where
$\mathbf{T}=\mathbf{A}\mathbf{\Psi}^{-1}\mathbf{H}^{H}\mathbf{H}\mathbf{A}^{H}$ and
$\bar{\mathbf{w}}=\mathbf{A}\mathbf{\Psi}^{-1}\mathbf{H}^{H}\mathbf{w}_{\mathrm T}$.
Based on \eqref{eq:DetectedSignal_ith}, the SINR} for the $i$-th chirp subcarrier can be expressed as \cite{MMSE_BER}
\begin{align}\label{eq:SINR}
    \beta_i = \frac{T(i,i)^2}{\mathrm{Var}\left(\sum_{j \neq i} T(i,j) x(j) + \bar{w}(i)\right)}=\frac{T(i,i)}{1 - T(i,i)}.
\end{align}
where $\mathrm{Var}(\cdot)$ denotes the variance operator.

The theoretical average BER of ZP-AFDM systems employing PSK or QAM modulation with Gray coding can be approximated as a function of SINR, given by\footnote{{It should be noted that the matrix \(\mathbf{T}\) and the average BER in \eqref{eq:MMSE_AverageBER} depend on the DAFT matrix \(\mathbf{A}\), which is uniquely determined by AFDM chirp parameters $c_1$ and $c_2$.}}
\begin{align}\label{eq:MMSE_AverageBER}
P_\mathrm{MMSE} 
&= \frac{1}{N} \sum_{i=0}^{N-1} a_M \operatorname{erfc}\left(\sqrt{\frac{b_M T(i,i)}{1 - T(i,i)}}\right),
\end{align}
where $a_M$ and $b_M$ are constants depending on the modulation scheme \cite{BER_Mapping}, and the complementary error function is defined as $\mathrm{erfc}(x) = \frac{2}{\sqrt{\pi}} \int_x^\infty e^{-t^2} dt$.

\section{Simulation Results}\label{Sec:Results}
In this section, we present simulation results to evaluate the performance of the proposed ZP-AFDM system and low-complexity detectors. {Unless otherwise specified}, the channel coefficients are modelled as \( h_i \sim \mathcal{CN}(0, 1/P) \), and the maximum delay spread is set as $Q = P-1$. The Doppler shift of each path is generated according to Jake's model, i.e., $\nu_i = \nu_{\max} \cos(\theta_i)$, where $\theta_i \sim \mathcal{U}[-\pi, \pi]$. For fair comparison, the ZP-AFDM system employs the same overall transmit power as the CPP-AFDM system.

\begin{figure}
    \centering
    \includegraphics[width=75mm]{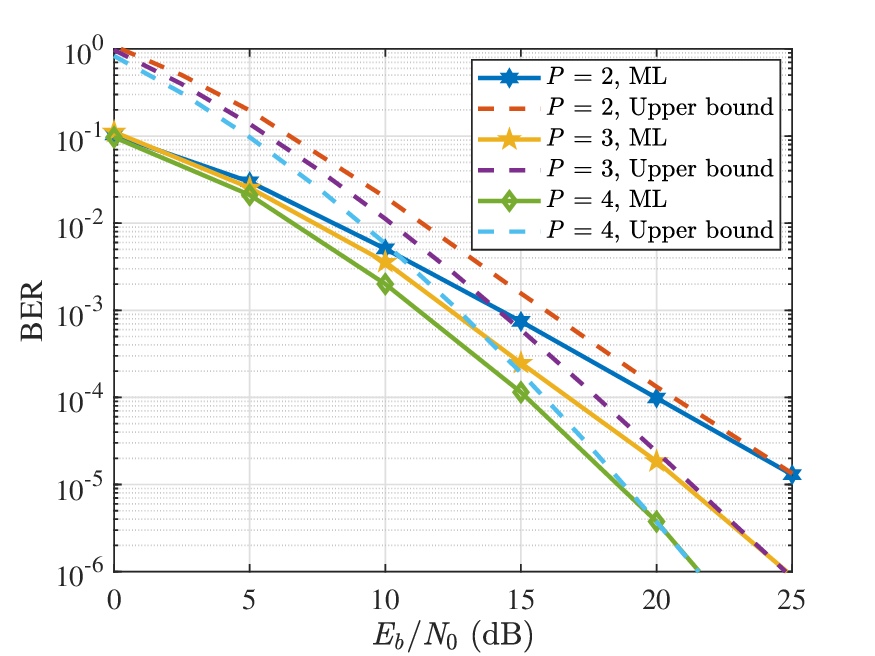}
    \caption{BER performance of the ML detector and upper bounds for the ZP-AFDM system with different numbers of paths.}
    \label{fig:ZPAfdm_ML_UpperBound}
\end{figure}

Fig.~\ref{fig:ZPAfdm_ML_UpperBound} shows the BER performance of the ML detector and its theoretical upper bounds for {the} ZP-AFDM system with $N = 6$, $N_\mathrm{ZP} = 2$, and BPSK modulation. {Moreover, doubly selective channels with $P=\left \{2,3,4\right \} $ paths are considered, and the normalized maximum Doppler shift is set to $\nu_{\max} = 1$.} It can be seen that the BER {curves of ZP-AFDM closely approach the analytical upper bounds} at high SNRs, validating the effectiveness of our derived error performance analysis. In addition, increasing $P$ leads to improved BER performance due to enhanced multipath diversity. 

\begin{figure}
    \centering
    \includegraphics[width=74mm]{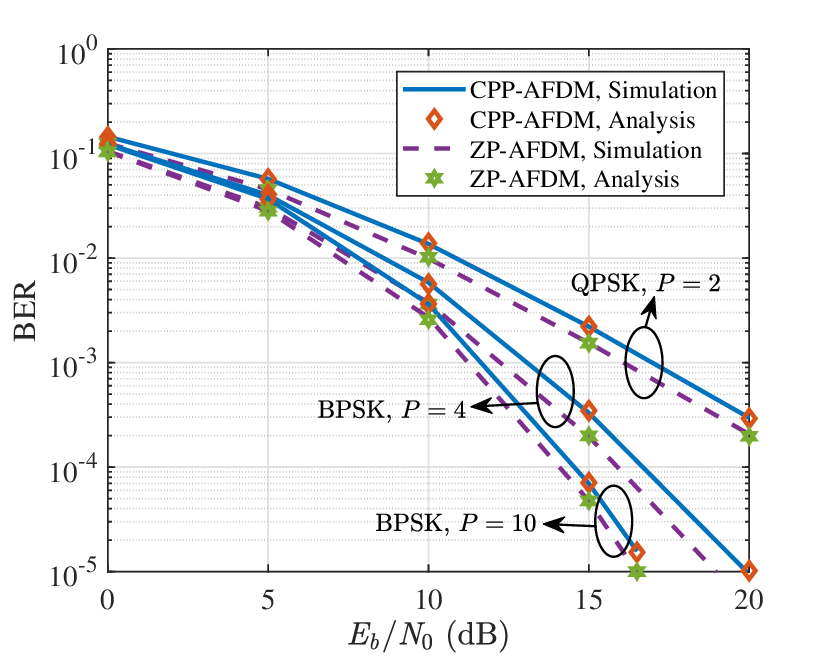}
    \caption{{Simulated and analytical BER comparisons of CPP-AFDM and ZP-AFDM systems using conventional MMSE detector under $\nu_{\max} = 0.5$.}}
 \label{fig:BER_ZPAfdmVsCPAfdm_MMSE}
\end{figure}

Next, we characterize the BER performance of the conventional MMSE and the proposed low-complexity detectors. Unless otherwise specified, all subsequent simulations are conducted using $N = 256$, $N_\mathrm{ZP} = 64$ \cite{ZP-OFDM}.
{Fig.~\ref{fig:BER_ZPAfdmVsCPAfdm_MMSE} compares the simulated and analytical BER of CPP-AFDM and ZP-AFDM systems using the conventional MMSE detector under three configurations: ``BPSK, $P=10$'', ``BPSK, $P=4$'', and ``QPSK, $P=2$''.}
As shown in Fig.~\ref{fig:BER_ZPAfdmVsCPAfdm_MMSE}, there is a perfect overlap between the analytical curves and the simulated results, demonstrating the accuracy of our derived BER expressions {for} the MMSE detector. Moreover, ZP-AFDM consistently outperforms CPP-AFDM, {as it allocates all transmit energy to data symbols, whereas CPP-AFDM needs to allocate the entire} transmit power across both data and prefixes. This trend is consistent with the observation in \cite{ZP-OFDM}, where ZP-OFDM outperforms CP-OFDM in terms of BER.

\begin{figure}
    \centering
    \includegraphics[width=75mm]{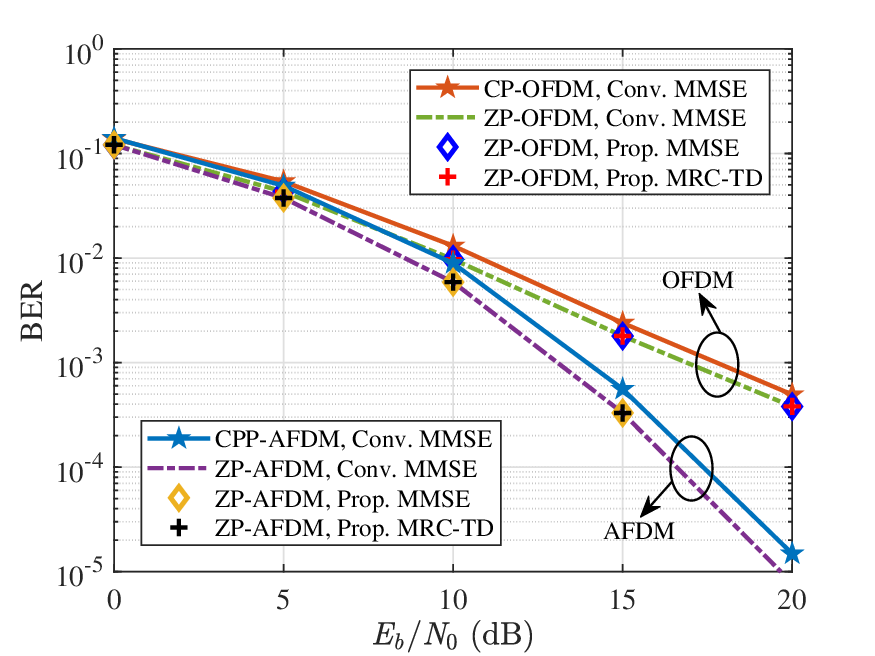}
    \caption{ 
    {BER performance of ZP-AFDM and ZP-OFDM employing the conventional MMSE, proposed low-complexity MMSE, and proposed MRC-TD detectors, as well as CPP-AFDM and CP-OFDM using the conventional MMSE detector under the EVA channel model \cite{EVA_Channel_R1Add}.}}
\label{fig:BER_AfdmVsOfdm_vmax1}
\end{figure}

Fig.~\ref{fig:BER_AfdmVsOfdm_vmax1} illustrates the BER performance of {ZP-AFDM and ZP-OFDM systems employing the conventional MMSE, proposed low-complexity MMSE, and proposed MRC-TD detectors, as well as CPP-AFDM and CP-OFDM systems using the conventional MMSE detector.} {The simulations are conducted with QPSK modulation under the Extended Vehicular A (EVA) channel model \cite{EVA_Channel_R1Add}.} It can be observed that the proposed low-complexity MMSE detector {for ZP-AFDM achieves the same BER as its conventional counterpart, while a similar trend can be observed in ZP-OFDM.} In addition, {the proposed MRC-TD detector for ZP-AFDM is capable of achieving BER performance nearly identical to that of the conventional MMSE detector.} Moreover, both ZP-AFDM and CPP-AFDM significantly outperform their OFDM counterparts.
{In particular, at a BER of \(10^{-3}\), ZP-AFDM and CPP-AFDM provide approximately \(E_b/N_0\) gains of 3.81 dB and 3.82 dB over ZP-OFDM and CP-OFDM, respectively.}

\begin{figure}
    \centering
    \begin{subfigure}{0.5\textwidth} 
        \centering
        \includegraphics[width=75mm]{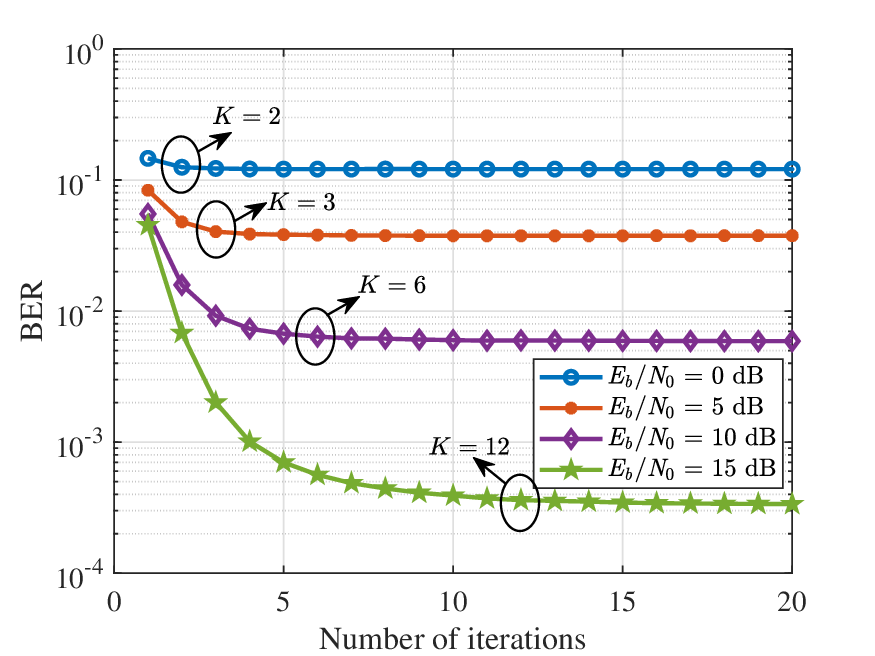}
        \subcaption*{(a)}
        \label{fig:MRCTD_Convergence}
    \end{subfigure}
    \hfill
    \begin{subfigure}{0.5\textwidth}
        \centering
        \includegraphics[width=75mm]{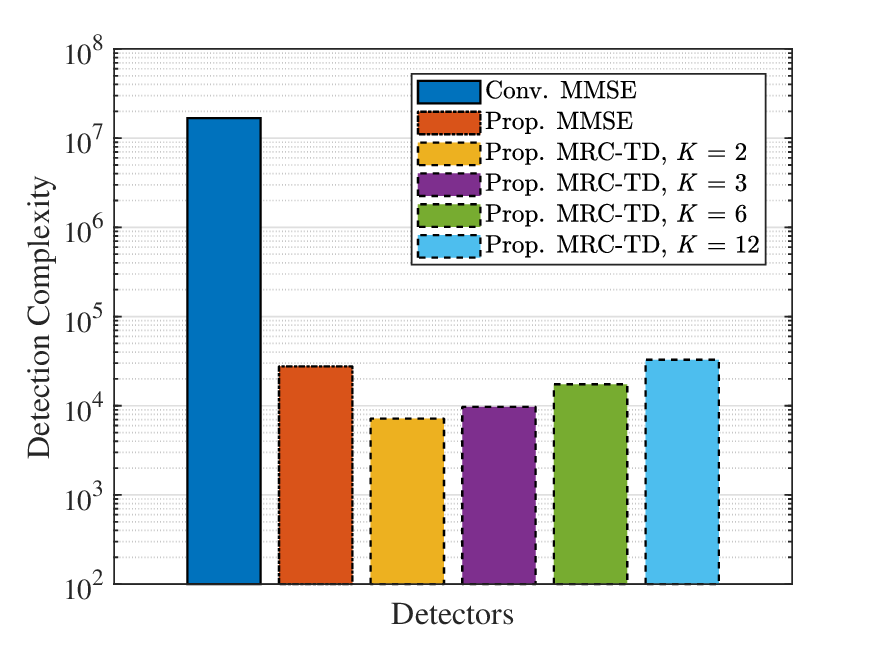}
        \subcaption*{(b)}
        \label{fig:Complexity_DiffDetection}
    \end{subfigure}
    \caption{{Convergence behavior of the proposed iterative MRC-TD detector and detection complexity comparison for ZP-AFDM: (a) BER convergence of the proposed MRC-TD detector versus the number of iterations under different $E_b/N_0$ values; (b) complexity comparison of the conventional MMSE, proposed low-complexity MMSE, and proposed MRC-TD detectors, 
    where the number of iterations for the MRC-TD detector is set to the converged values obtained from Fig.~7(a), i.e., $K=\{2,3,6,12\}$ for $E_b/N_0=\{0,5,10,15\}$~dB, respectively.}}
    \label{fig:Convergence_Complexity}
    \vspace{-1.5em}
\end{figure}

{In Fig.~\ref{fig:Convergence_Complexity}, we investigate the convergence behavior of the proposed iterative MRC-TD detector and the associated detection complexity in ZP-AFDM systems. The simulation parameters and channel settings are identical to those in Fig.~\ref{fig:BER_AfdmVsOfdm_vmax1}. Fig.~\ref{fig:Convergence_Complexity}(a) depicts the BER performance of the proposed MRC-TD detector versus the number of iterations under different $E_b/N_0$ values. 
It can be observed that increasing the number of iterations improves the detection performance, as the receiver can progressively mitigate residual interference and the BER converges after several iterations. Accordingly, the converged iteration numbers can be approximately summarized as $K=\{2,3,6,12\}$ for $E_b/N_0=\{0,5,10,15\}$~dB, respectively. Based on these convergence results, Fig.~\ref{fig:Convergence_Complexity}(b) compares the complexity of the conventional MMSE, proposed low-complexity MMSE, and proposed MRC-TD detectors, where the number of iterations for the MRC-TD detector is set to $K=\{2,3,6,12\}$. It can be seen that both the proposed low-complexity MMSE and MRC-TD detectors provide substantial complexity reductions while achieving BER performance nearly identical to that of the conventional MMSE detector.
Moreover, compared with the proposed low-complexity MMSE detector, MRC-TD offers a flexible BER vs. complexity trade-off by varying the number of iterations. }




\vspace{-1em}
\section{Conclusion}\label{Sec:Conclusion}
In this letter, we have proposed a ZP-AFDM system and derived its corresponding input-output relationship in the affine frequency domain. {By exploiting the unique lower-triangular and sparse structure of the {TD} channel matrix, low-complexity MMSE and MRC-TD detectors {have been} developed. Furthermore, the error rate performance of the ZP-AFDM system {has been derived} under both ML and MMSE detection. {Our numerical experiments have shown that 1) there is a perfect overlap between the derived BER and Monte-Carlo simulations; and 2) the proposed detectors can achieve {BER} performance nearly identical to that of {the} conventional MMSE detector with a substantial reduction in complexity.}

\end{document}